\documentclass[dvips]{article}
\usepackage{supertabular,lscape,epsfig}
\usepackage{amssymb}
\usepackage{amsmath}

\DeclareSymbolFont{ppa}{OT1}{ppl}{m}{it}
\DeclareMathSymbol{\vv}{\mathalpha}{ppa}{'166}

\begin{document}

\newcommand{\dd}{\,{\rm d}}
\newcommand{\ie}{{\it i.e.},\,}
\newcommand{\etal}{{\it et al.\ }}
\newcommand{\eg}{{\it e.g.},\,}
\newcommand{\cf}{{\it cf.\ }}
\newcommand{\vs}{{\it vs.\ }}
\newcommand{\zdot}{\makebox[0pt][l]{.}}
\newcommand{\up}[1]{\ifmmode^{\rm #1}\else$^{\rm #1}$\fi}
\newcommand{\dn}[1]{\ifmmode_{\rm #1}\else$_{\rm #1}$\fi}
\newcommand{\upd}{\up{d}}
\newcommand{\uph}{\up{h}}
\newcommand{\upm}{\up{m}}
\newcommand{\ups}{\up{s}}
\newcommand{\arcd}{\ifmmode^{\circ}\else$^{\circ}$\fi}
\newcommand{\arcm}{\ifmmode{'}\else$'$\fi}
\newcommand{\arcs}{\ifmmode{''}\else$''$\fi}
\newcommand{\MS}{{\rm M}\ifmmode_{\odot}\else$_{\odot}$\fi}
\newcommand{\RS}{{\rm R}\ifmmode_{\odot}\else$_{\odot}$\fi}
\newcommand{\LS}{{\rm L}\ifmmode_{\odot}\else$_{\odot}$\fi}

\newcommand{\Abstract}[2]{{\footnotesize\begin{center}ABSTRACT\end{center}
\vspace{1mm}\par#1\par
\noindent
{~}{\it #2}}}

\newcommand{\TabCap}[2]{\begin{center}\parbox[t]{#1}{\begin{center}
  \small {\spaceskip 2pt plus 1pt minus 1pt T a b l e}
  \refstepcounter{table}\thetable \\[2mm]
  \footnotesize #2 \end{center}}\end{center}}

\newcommand{\TableSep}[2]{\begin{table}[p]\vspace{#1}
\TabCap{#2}\end{table}}

\newcommand{\FigCap}[1]{\footnotesize\par\noindent Fig.\  %
  \refstepcounter{figure}\thefigure. #1\par}

\newcommand{\TableFont}{\footnotesize}
\newcommand{\TableFontIt}{\ttit}
\newcommand{\SetTableFont}[1]{\renewcommand{\TableFont}{#1}}

\newcommand{\MakeTable}[4]{\begin{table}[htb]\TabCap{#2}{#3}
  \begin{center} \TableFont \begin{tabular}{#1} #4 
  \end{tabular}\end{center}\end{table}}

\newcommand{\MakeTableSep}[4]{\begin{table}[p]\TabCap{#2}{#3}
  \begin{center} \TableFont \begin{tabular}{#1} #4 
  \end{tabular}\end{center}\end{table}}

\newenvironment{references}%
{
\footnotesize \frenchspacing
\renewcommand{\thesection}{}
\renewcommand{\in}{{\rm in }}
\renewcommand{\AA}{Astron.\ Astrophys.}
\newcommand{\AAS}{Astron.~Astrophys.~Suppl.~Ser.}
\newcommand{\ApJ}{Astrophys.\ J.}
\newcommand{\ApJS}{Astrophys.\ J.~Suppl.~Ser.}
\newcommand{\ApJL}{Astrophys.\ J.~Letters}
\newcommand{\AJ}{Astron.\ J.}
\newcommand{\IBVS}{IBVS}
\newcommand{\PASP}{P.A.S.P.}
\newcommand{\Acta}{Acta Astron.}
\newcommand{\MNRAS}{MNRAS}
\renewcommand{\and}{{\rm and }}
\section{{\rm REFERENCES}}
\sloppy \hyphenpenalty10000
\begin{list}{}{\leftmargin1cm\listparindent-1cm
\itemindent\listparindent\parsep0pt\itemsep0pt}}%
{\end{list}\vspace{2mm}}

\def\TYLDA{~}
\newlength{\DW}
\settowidth{\DW}{0}
\newcommand{\dw}{\hspace{\DW}}

\newcommand{\refitem}[5]{\item[]{#1} #2%
\def\REFARG{#3}\ifx\REFARG\TYLDA\else, {\it#3}\fi
\def\REFARG{#4}\ifx\REFARG\TYLDA\else, {\bf#4}\fi
\def\REFARG{#5}\ifx\REFARG\TYLDA\else, {#5}\fi.}

\newcommand{\Section}[1]{\section{#1}}
\newcommand{\Subsection}[1]{\subsection{#1}}
\newcommand{\Acknow}[1]{\par\vspace{5mm}{\bf Acknowledgements.} #1}
\pagestyle{myheadings}

\newfont{\bb}{ptmbi8t at 12pt}
\newcommand{\xrule}{\rule{0pt}{2.5ex}}
\newcommand{\xxrule}{\rule[-1.8ex]{0pt}{4.5ex}}
\def\thefootnote{\fnsymbol{footnote}}
\begin{center}
{\Large\bf Comparison of Parallaxes from Eclipsing Binaries Method with
Hipparcos Parallaxes}

\vskip1cm
{\bf Irena~~~~S~e~m~e~n~i~u~k}
\vskip3mm
{Warsaw University Observatory, Al.~Ujazdowskie~4, 00-478~Warszawa,
Poland\\
e-mail: is@astrouw.edu.pl}
\end{center}

\vskip10pt
\Abstract{The parallaxes determined by Lacy (1979) by means of eclipsing 
binaries method are compared with the Hipparcos parallaxes for 19 systems. The 
residual scatter of the distance moduli inferred from eclipsing binaries 
method -- after allowing for known errors as given by Lacy and Hipparcos -- is 
equal to 0.18~mag. It decreases to 0.08~mag when obviously not fitting 
semi-detached systems and systems with chromospheric activity of components 
are removed from the sample.}{binaries: eclipsing -- Stars: distances} 

\vskip1.5cm
The parallax determination by means of double-line eclipsing binaries is now 
considered to be one of the most promising methods of distance determination 
(\eg Pa\-czy{\'n}ski 1997, see also Kruszewski and Semeniuk 1999 for a 
historical review). However, before using eclipsing binaries as unquestionable 
standard candles there is a need to check the accuracy of distances obtained 
in this way by comparing them with the distances determined with other 
methods, particularly with the distances from trigonometric parallaxes. The 
{\it Hipparcos Catalogue} (ESA 1997) provides us with trigonometric parallaxes 
for several hundred eclipsing binaries. Popper (1998) used the Hipparcos data 
as a check of the  eclipsing binaries method. Based on known solutions of 
photometric and spectroscopic orbits he calculated surface brightness of 14 
detached eclipsing binaries closer than 125~pc and with the Hipparcos parallax 
error not greater than 10\%. The sample was generally not homogeneous. Six of 
the fourteen systems had chromospherically active or intrinsically variable 
components. The ${(B-V)}$ color indices were taken from different sources and 
some were uncertain. Popper compared the surface brightnesses of these 14 
stars with the calibrated by him (Popper 1980) relation between surface 
brightness and ${(B-V)}$ color index. The comparison indicated some 
deviations from this relation. In particular the early type stars, with ${(B-
V)}$ less than 0.04, were located slightly above the relation, while almost 
all chromospherically active later type stars were situated beneath the 
relation, generally more than the typical uncertainty of the surface 
brightness determination. As suggested by Popper the surface brightnesses of 
the latter stars were depressed due to spots on their surfaces. The Popper 
general conclusion was that the relation between surface-brightness and ${(B-
V)}$ color index for the stars of lower temperature was poorly established. 
Recently Oblak and Kurpi{\'n}ska-Winiarska (2000) compared the Hipparcos 
parallaxes with photometric parallaxes calculated for the eclipsing binaries 
by Brancewicz and Dworak (1980), and revised by Jacob (1999). They were 
aware of crudeness of the photometric parallaxes obtained from the 
mass-luminosity relation instead of double-lined spectroscopic orbits, as well 
as of inhomogeneity of their rich (338 systems in total) sample. They found 
that for the majority of Brancewicz and Dworak (1980) eclipsing binaries the 
photometric parallaxes were comparable with Hipparcos parallaxes and that for 
103 stars with well determined Hipparcos parallaxes the mean error of the 
absolute magnitude differences derived from Hipparcos and given by Brancewicz 
and Dworak is less than 1~mag. The paper showed that the photometric 
parallaxes as determined by Brancewicz and Dworak could be used as approximate 
distance determination but are not good for accurate distance determination. 
There is, however, in the literature a sample which seems to be more suitable 
for verification of usefulness of the eclipsing binaries method for distance 
determination than the samples of Popper or of Dworak and Brancewicz. This is 
a homogeneous sample prepared by Lacy (1979). 

Lacy (1979) using the Barnes--Evans (1976) method determined distance moduli 
${m-M=V_0-M_V}$ for 47 eclipsing binaries with known absolute dimensions based 
on double-lined spectroscopic orbits. The relevant equation he used was 
$$M_V=42.362-5\log R/\RS-10 F_V(V-R)\eqno(1)$$ 
where the visual surface brightness parameter ${F_V(V-R)}$ was taken from the 
surface brightness -- color relation as given by Barnes, Evans and Parsons 
(1976). The homogeneity of the Lacy's sample is secured by homogeneously 
determined -- from the author's observations -- ${(V-R)}$ color indices and 
interstellar reddening. This is important because systematic errors in colors 
are dangerous, especially when they are different for different objects. 
                        
To check the accuracy of this method of distance determination we have 
compared the distances determined by Lacy with the distances obtained by the 
trigonometric parallax method from the {\it Hipparcos Catalogue} (ESA 1997). 
>From the Lacy's sample of 47 eclipsing binaries we selected 19 EA-type 
systems with the Hipparcos parallax error less than 20\%. All but one (CM~Lac) 
stars turned out to be closer than 200~pc. These stars are listed in Table~1. 
The variability and the spectral types in columns 3 and 4 of the Table are 
taken from the {\it Hipparcos Catalogue}. The dereddened ${(V-R)_0}$ colors in 
column~5 are calculated from Lacy's (1979) data. Column~6 contains the 
difference ${\Delta(m-M)}$ of the Lacy's and Hipparcos distance moduli. Columns 
7, 8, 9 and 10 give the Lacy's and Hipparcos parallaxes and their standard 
errors. All the values are in milliarcseconds (mas). The Lacy's parallax errors 
in column~8 were calculated from the formula  
$$\sigma_{m-M}=5{\sigma_\pi\over\pi}\log_{10}e\eqno(2)$$   
where the distance modulus error ${\sigma_{m-M}}$ was taken to be 0.15~mag for 
OB spectral type stars and 0.11 mag for later spectral type stars, as given by 
Lacy (1979).  
\MakeTable{l@{\hspace{-3pt}}rll@{\hspace{-5pt}}rrrrrr}{12.5cm}{Nearby Eclipsing Binaries with parallaxes 
determined by Lacy and Hipparcos}
{\hline
\noalign{\vskip3pt}
\multicolumn{1}{c}{Name}&
\multicolumn{1}{c}{HIP}&
\multicolumn{1}{c}{Var}&
\multicolumn{1}{c}{Spectral}&
\multicolumn{1}{c}{$(V{-}R)_0$}&
\multicolumn{1}{c}{$\Delta(m{-}M)$}&
\multicolumn{1}{c}{$\pi_{\rm Lacy}$}&
\multicolumn{1}{c}{$\sigma_{\rm Lacy}$}&
\multicolumn{1}{c}{$\pi_{\rm Hip}$}&
\multicolumn{1}{c}{$\sigma_{\rm Hip}$}\\
&
\multicolumn{1}{c}{Number}&
\multicolumn{1}{c}{type}&
\multicolumn{1}{c}{type}&&&
\multicolumn{1}{c}{[mas]}&
\multicolumn{1}{c}{[mas]}&
\multicolumn{1}{c}{[mas]}&
\multicolumn{1}{c}{[mas]}\\
\noalign{\vskip3pt}
\hline
WW Aur      &  31173 & EA/DM & A3m+A3m &  0.146  &  0.10~~~& 11.32 &  0.57 & 11.86 & 1.06\\
AR Aur      &  24740 & EA    & B9.5V   &  0.009  &  0.23~~~&  7.38 &  0.37 &  8.20 & 0.78\\
$\beta$ Aur &  28360 & EA    & A2V     &  0.077  &--0.34~~~& 46.56 &  2.36 & 39.72 & 0.78\\
ZZ Boo      &  68064 & EA/DM & F2V     &  0.339  &  0.10~~~&  8.47 &  0.43 &  8.88 & 0.78\\
EI Cep      & 106024 & EA/DM & F2V     &  0.200  &  0.09~~~&  4.83 &  0.24 &  5.03 & 0.56\\
V1143 Cyg   &  96620 & EA/DM & F6Vasv  &  0.407  &  0.08~~~& 24.21 &  1.23 & 25.12 & 0.56\\
Z Her       &  87965 & EA/AR & F6V     &  0.540  &--0.59~~~& 13.37 &  0.68 & 10.17 & 0.84\\
TX Her      &  84670 & EA/DM & A9V     &  0.269  &--0.14~~~&  5.92 &  0.30 &  5.55 & 0.84\\
V624 Her    &  86809 & EA    & A3m     &  0.138  &--0.12~~~&  7.31 &  0.37 &  6.93 & 0.74\\
HS Hya      &  50966 & EA/D  & F5V     &  0.414  &  0.42~~~&  9.08 &  0.46 & 11.04 & 0.88\\
CM Lac      & 108606 & EA/DM & A2V     &  0.163  &--0.45~~~&  5.42 &  0.27 &  4.40 & 0.84\\
UV Leo      &  52066 & EA/DW & G0V     &  0.579  &--0.20~~~& 11.91 &  0.60 & 10.85 & 1.16\\
$\delta$ Lib&  73473 & EA/SD & B9.5V   &  0.080  &  0.67~~~&  7.87 &  0.40 & 10.72 & 0.91\\
RR Lyn      &  30651 & EA/DM & A3m     &  0.192  &--0.02~~~& 12.13 &  0.61 & 12.01 & 0.97\\
U Oph       &  84500 & EA/DM & B5Vnn   &--0.086  &  0.63~~~&  4.02 &  0.20 &  5.38 & 0.83\\
WZ Oph      &  83719 & EA/DM & F8V     &  0.499  &  0.34~~~&  6.82 &  0.35 &  7.99 & 1.37\\
$\beta$ Per &  14576 & EA/SD & B8V     &  0.016  &  0.17~~~& 32.51 &  1.65 & 35.14 & 0.90\\
CD Tau      &  24663 & EA/D  & F7V     &  0.431  &  0.04~~~& 13.43 &  0.68 & 13.66 & 1.64\\
BH Vir      &  68258 & EA/DW & F8V     &  0.543  &  0.14~~~&  7.45 &  0.68 &  7.94 & 1.50\\
\hline}

Fig.~1 compares Hipparcos parallaxes \vs Lacy's parallaxes. The Lacy's and 
Hipparcos parallax errors are also plotted. The most deviating star is 
$\beta$~Aur, which is also the star with the largest parallax.  
\begin{figure}[htb]
\centerline{\includegraphics[width=12.7cm]{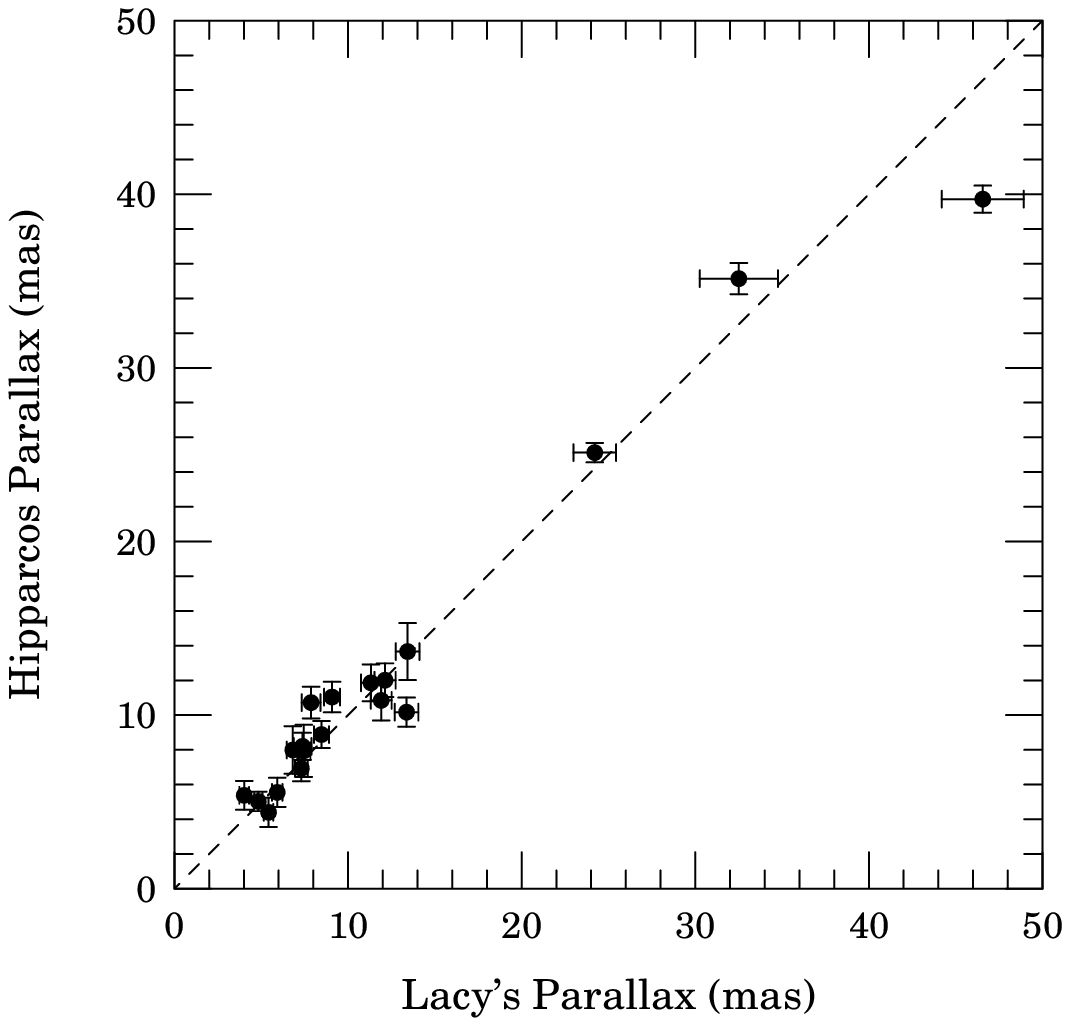}}
\FigCap{The Hipparcos parallaxes plotted \vs Lacy's parallaxes together with 
their standard errors. The closest and the most deviating star is 
$\beta$~Aur.} 
\end{figure}

Based on the values of column 6 we have calculated the value of variance for 
the difference of Lacy's and Hipparcos distance moduli. This value is equal to 
0.1093 what gives 0.33~mag for the standard deviation of this difference. 
Having the Hipparcos parallax errors (column~10) we could estimate how much 
they contribute to this value. The errors of the Hipparcos distance moduli of 
individual stars, calculated with Eq.~(2), give 0.0623 for the corresponding 
variance, so the variance resulting from the Lacy's distance moduli only 
is 0.0470, what gives for the standard deviation the value equal to 0.22~mag. 
The situation improves significantly if we reject from our table the 
semi-detached systems and the stars with the chromospheric activity of 
components. There are five such systems in our sample, two semi-detached 
systems ($\delta$~Lib, $\beta$~Per) and three binaries with chromospherically 
active components (Z~Her, UV~Leo, BH~Vir). After rejecting these stars we 
obtain 0.14~mag as the standard deviation corresponding to the Lacy's distance 
moduli for the 14 remaining systems. With the observational standard errors of 
the Lacy's distance moduli (0.15~mag for OB stars and 0.11~mag for later 
types), dominated by errors in absolute dimensions of the stars, we can remove 
their contribution from the corresponding variance of the distance moduli. The 
resulting difference gives 0.08~mag, as a residual scatter, for the 14 
detached systems. For the full sample of 19 systems this residual scatter is 
equal to 0.18~mag. 

\begin{figure}[htb]    
\centerline{\includegraphics[width=12.5cm]{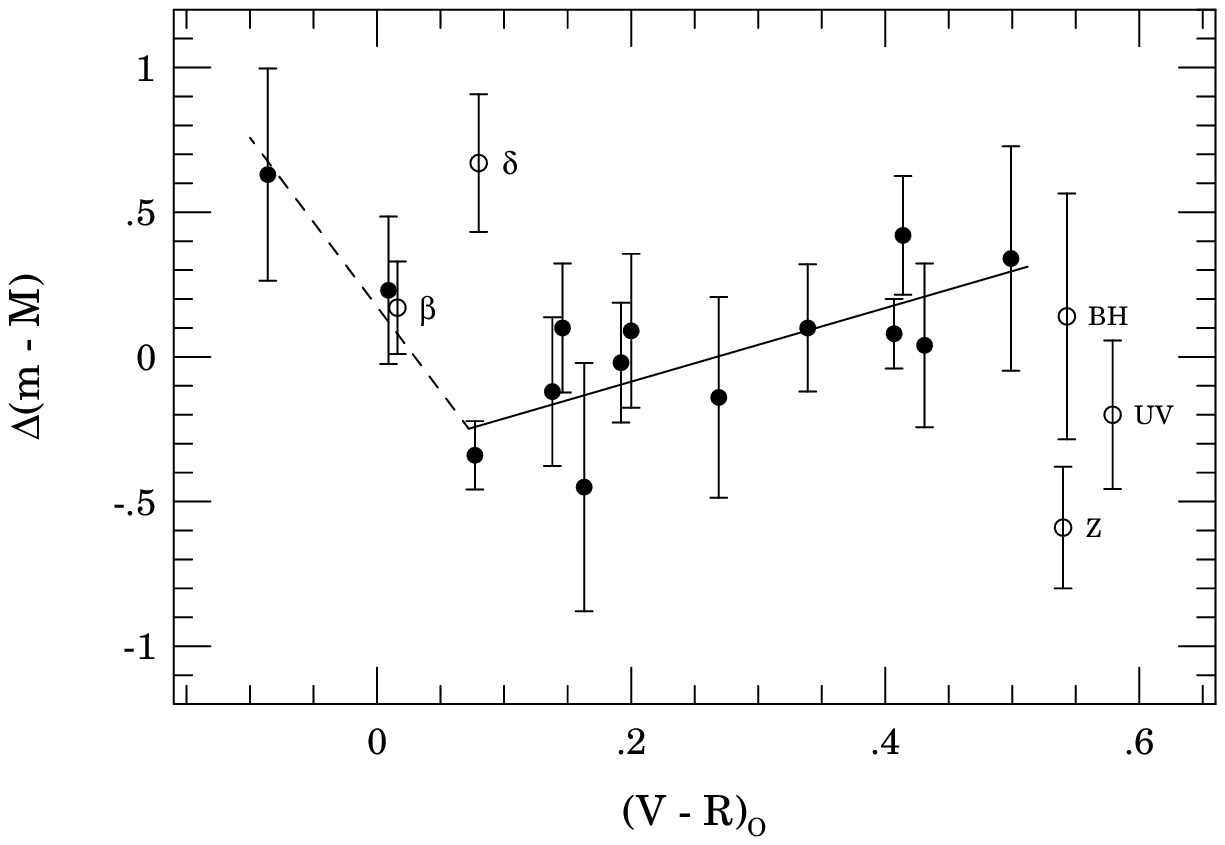}} \FigCap{The relation 
between the difference of the Lacy's and Hipparcos distance moduli and the 
color index ${(V-R)_0}$. The error bars correspond to the total error of the 
difference resulting from the Lacy's and Hipparcos standard errors. The dots 
denote the detached systems without observed chromospheric activity, the open 
circles correspond to the semi-detached systems ($\delta$~Lib and $\beta$~Per) 
or systems with a chromospheric activity (Z~Her, UV~Leo and BH~Vir). The line 
segments plotted in the figure are obtained with the least square method for 
the dots only.} 
\end{figure}
Fig.~2 gives the dependence of the Lacy's and Hipparcos distance moduli 
difference on the color index ${(V-R)_0}$, together with the total error of 
the difference resulting both from the Hipparcos and Lacy's determination 
errors. The open circles denote the semi-detached systems and the 
chromospherically active stars. If we confine our consideration to the 
detached systems without chromospheric activity (dots) we see that the 
dependence between the distance modulus difference and the color index could 
be described by two linear relations with different slopes. The lines in  
Fig.~2 were obtained with the least square method. The slope changes at 
approximately ${(V-R)_0=0.07}$. For the stars with redder colors the distance 
moduli difference grows with the color what means that the Lacy's distance 
modulus becomes greater than the Hipparcos distance modulus. The opposite 
seems to be true for the stars with colors less than 0.07~mag. We suggest that 
this change of slope in the relation presented in Fig.~2 is related to the 
change of slope in the relation between the surface brightness parameter $F_V$ 
and ${(V-R)_0}$ as given by Barnes, Evans and Parsons (1976). It should 
be mentioned here that all the standard deviations calculated in the preceding 
paragraph do not take into account this change of slope of the relation 
visible in Fig.~2. 
 
The star that lies almost exactly on the intersection of the two line segments 
in Fig.~2 is $\beta$~Aur. Perhaps this circumstance  explains partly its most 
deviating location in Fig.~1. Also the fact that $\beta$~Aur has very shallow 
eclipses -- their depths are equal to only 0.08~mag -- could explain the 
apparent deviation. 

In conclusion we can state that the test performed on the data of the Lacy's 
(1979) sample does not contradict the usefulness of the eclipsing binaries 
method for the distance determination. The results of our analysis are 
consistent with the conjecture that the residual scatter of the distance 
modulus obtained in this way -- after allowing both for the Lacy's and 
Hipparcos observational errors -- is less than 0.1~mag provided that we reject 
semi-detached and chromospherically active systems. This rejection could be 
made without difficulties, as we have precise criteria for selecting these 
stars. These are the variability between minima for the chromospherically 
active systems and the results of photometric orbit solution for the 
semi-detached systems. It is hoped that this residual scatter will be reduced 
in the future as the observational errors are reduced, and the modeling of 
eclipsing binaries improves. Here we should remind, however, that the Lacy's 
sample contains the stars from the Sun vicinity with homogeneous population 
features. We cannot exclude that the application of this method to the systems 
with different population characteristics would increase the scatter.   
 
\Acknow{The author is greatly indebted to Prof.~Andrzej Kru\-szew\-ski for 
suggesting the subject of this paper and for many helpful discussions. Special 
thanks are also due to Professor Bohdan Paczy{\'n}ski for reading and 
commenting on the manuscript. This work would not be possible without use of 
the {\it Hipparcos Catalogue}. A partial support from the KBN grant BST to the 
Warsaw University Observatory is also acknowledged.}

\end{document}